\documentclass[epj]{svjour}

\usepackage{graphicx}
\usepackage{fancyhdr}
\usepackage{bm}
\def\makeheadbox{{
 }}
\newcommand{\nl}{\nonumber \\ }

\newcommand{\be}{\begin{equation}}  
\newcommand{\ee}{\end{equation}}  
\newcommand{\bea}{\begin{eqnarray}}  
\newcommand{\eea}{\end{eqnarray}}  

\def\slash#1{#1\!\!\!/\!\,\,}

\def\mytitle{My title} 
\def\myauthors{My name}  
\def\mytype{My type of session}
\def\mysession{My session}

\def\mytitle{Topological physics} 
\def\myauthors{Richard J. Hill}    
\def\mytype{Contributed Talk}    
\def\mysession{Alternatives}

\begin{document}
\title{Topological physics in the standard model and beyond}
\author{Richard J. Hill\inst{}
}                     
%
%
\institute{
Fermi National Accelerator Laboratory\\
\it P.O. Box 500, Batavia, Illinois 60510, USA
}
%
\date{}
\abstract{
Topological interactions are an 
essential ingredient for building consistent low-energy theories of fermions, gauge 
fields and Nambu-Goldstone bosons in the
absence of explicit UV completions, such as in Little Higgs theories.   
These interactions are also a probe of UV completion physics that may be 
out of direct experimental reach. 
The technology of topological, or Wess-Zumino-Witten interactions 
is described, using explicit examples in the standard model
and in Little Higgs models.  
The construction of a simple topological action on 
$SU(3)/SU(2)$ is described. 
Inconsistencies in some popular Little Higgs models are pointed out. 
} 
\maketitle
\section{Introduction}
\label{intro}

Topological interactions and anomaly physics 
are of basic importance
in the standard model, 
in model-building applications beyond 
the standard model, 
and in formal studies of quantum field theory.  

However, the methods used to study these interactions are not part of the 
everyday theoretical toolkit.  
This may be due partly to the perceived complexity of the 
mathematics involved, 
and partly to the perceived scarcity of relevant physics 
applications.  
My goal in this talk is to help dispel these notions, and to point
out some new applications of these tools. 

Much of this perceived complexity 
is associated with ancient notions of 
current algebra that still pervade many discussions of anomalies. 
Section~\ref{sec:sm} reviews some simple processes in the 
standard model from a modern effective field theory point of 
view.  
To illustrate the mathematical simplicity of this physics, 
a new and especially simple topological construction for $SU(3)/SU(2)$ 
is described  in Section~\ref{sec:simplest};  
the most complicated mathematical object needed here is a sphere. 
Section~\ref{sec:lh} describes the application of these tools 
to Little Higgs models, pointing out some confusions regarding
gauge invariance, anomalies, and spurious parities that 
have afflicted the literature. 
Section~\ref{sec:conclusion} concludes by mentioning some more 
formal applications of the technology that was developed to study 
Little Higgs models. 

\section{Standard model examples} 
\label{sec:sm}

There are a number of applications, both new and old, of
topological interactions in the standard model.  We mention
three examples here.  

\subsection{$\pi^0\to\gamma\gamma$ and the QCD chiral lagrangian}

Textbook treatments~\cite{peskin} of the famous 
$\pi^0\to \gamma \gamma$ decay may lead the uninitiated 
reader to believe that computing anomalous divergences of
axial vector currents, 
and working through subtle regularization schemes,
are prerequisite to making rigorous predictions.  
In modern effective field theory language, the situation is much simpler.
Once we know the fields in our effective theory 
(a matrix field $U(x)$ taking values in 
$SU(n_f)\times SU(n_f)/SU(n_f)= SU(n_f)$, with $n_f$ the number of
quark flavors)
and the symmetries of our effective theory 
(global $SU(n_f)\times SU(n_f)$), 
we can write down the most general operator made from these fields, 
and obeying these symmetries.  
One such operator---the Wess Zumino Witten (WZW) term~\cite{Wess,Witten}---happens 
to have the remarkable property
that it uniquely predicts the $\pi^0\to\gamma \gamma$ rate.  
In detail, when expanded onto the relevant fields, this interaction 
reads~\cite{BJ,Adler,Bardeen}
\be
\Gamma_{WZW} = - {N_c\over 96\pi^2 f_\pi } \int d^4x\, \epsilon^{\mu\nu\rho\sigma} 
{\pi^0} F_{\mu\nu} \,  F_{\rho\sigma} \,,
\ee
where $N_c=3$ is the number of quark colors, and $f_\pi \approx 93\,{\rm MeV}$ 
is the pion decay constant. 
From this interaction Lagrangian, we can simply write down the 
appropriate Feynman diagram and compute the decay rate. 

\subsection{Weak currents in the QCD chiral lagrangian \label{sec:weak} } 

Although processes like
$\pi^0\to\gamma\gamma$ involving just pseudoscalars and photons 
provide  
the most familiar examples of gauged WZW terms, 
the low-energy standard 
model also contains the charged and neutral weak currents, 
vector mesons coupling to baryon number and isospin, 
and the hypothetical axion.
Incorporating these ingredients into the WZW structure leads to 
interesting effects.   
An example is the neutrino-photon interaction mediated by~\cite{Harvey:2007rd} 
\be\label{nugamma}
\Gamma_{WZW} = {N_c\over 48\pi^2}{ e g_\omega g_2 \over \cos\theta_W } 
\int d^4x\, \epsilon^{\mu\nu\rho\sigma} \omega_\mu Z_\nu F_{\rho\sigma} \,,
\ee
where $\omega$ is the isoscalar vector meson coupling to baryon
number.  The interaction (\ref{nugamma}) will lead to the detection mode
$\nu + N \to \nu + N + \gamma$ in laboratory neutrino experiments; 
and to the cooling mechanism $\gamma\to \nu\bar{\nu}$ in 
neutron stars.~\cite{Harvey:2007rd}.

\subsection{WZW for the standard model Higgs} 

As a prelude to applications in Little Higgs models,
it is instructive to consider the WZW term for the Higgs boson in the
standard model.  
Consider, for example, the situation where the third generation 
$(t,b)$ quarks are
integrated out, leaving an effective Lagrangian 
involving just the $(\nu_\tau, \tau)$ leptons, 
and the complete set of first and second generation fermions.   
We would like to think that the result is a consistent and predictive 
effective field theory. 
In particular, the low-energy 
theory should be gauge invariant under $SU(2)_L\times U(1)_Y$.   
Naively, it appears that such an effective theory is not possible---an 
inspection of triangle diagrams for the remaining fermions shows that there
are uncancelled gauge anomalies from the third-generation leptons. 

Of course, the resolution to this paradox is clear---we've 
left out an operator.
At low energies, the only other matter fields 
in the theory besides the fermions 
are the Goldstone modes of the Higgs, represented by
a field $\phi$ in $SU(2)\times U(1)/U(1) \cong S^3$ ($S^3$ is the 
three-sphere): 
\be
\phi = \exp\left[{i\over v} \left(\begin{array}{cc} 0 & g^+ \\ g^- & g^0 \end{array}\right) \right]
\left(\begin{array}{c} 0 \\ v \end{array} \right) \,.
\ee
This field transforms as an electroweak doublet:
\be
\delta \phi =i( \epsilon +  \epsilon') \phi \,,
\ee
where $\epsilon$ and $\epsilon'$ generate $SU(2)_L$ and $U(1)_Y$.    
What operators can we build out of $\phi$ ?  

It turns out that up to normalization, 
there is an essentially unique operator
that can be  built out of $\phi$ and the gauge bosons $W$, $B$
that: 
(i) is globally invariant under $SU(2)\times U(1)$; 
and (ii) generates a consistent anomaly in the spontaneously broken 
generators.  
As outlined in Section~\ref{sec:simplest}, 
the explicit form of this operator can be obtained from a topological 
construction.  
Explicit calculation shows that 
under a general $SU(2)\times U(1)$ gauge transformation, 
the operator produces an anomaly proportional to the uncancelled lepton 
anomaly. 
Enforcing gauge invariance (that is, anomaly cancellation) 
then fixes the overall 
normalization factor.~\cite{D'Hoker:1984pc}   

This example illustrates an important point in building consistent 
models of fermions, gauge fields and pNGB's,  
for example Little Higgs models.  
In order to enforce gauge invariance, we must
be able to find operators made out of the fields in our theory that cancel
any left-over gauge anomalies from explicit fermion degrees of freedom. 
Such a construction is guaranteed in the present case---we started with the consistent standard model, and by integrating
out heavy degrees of freedom we should end up with a 
consistent low-energy theory.  
In bottom-up approaches where we don't know the
UV theory, the possibility of anomaly cancellation is not 
guaranteed.  Additional degrees of freedom might be necessary for consistency. 

\section{The simplest WZW term}
\label{sec:simplest}

\begin{figure}
\begin{center}
\includegraphics[width=0.2\textwidth,height=0.2\textwidth,angle=0]{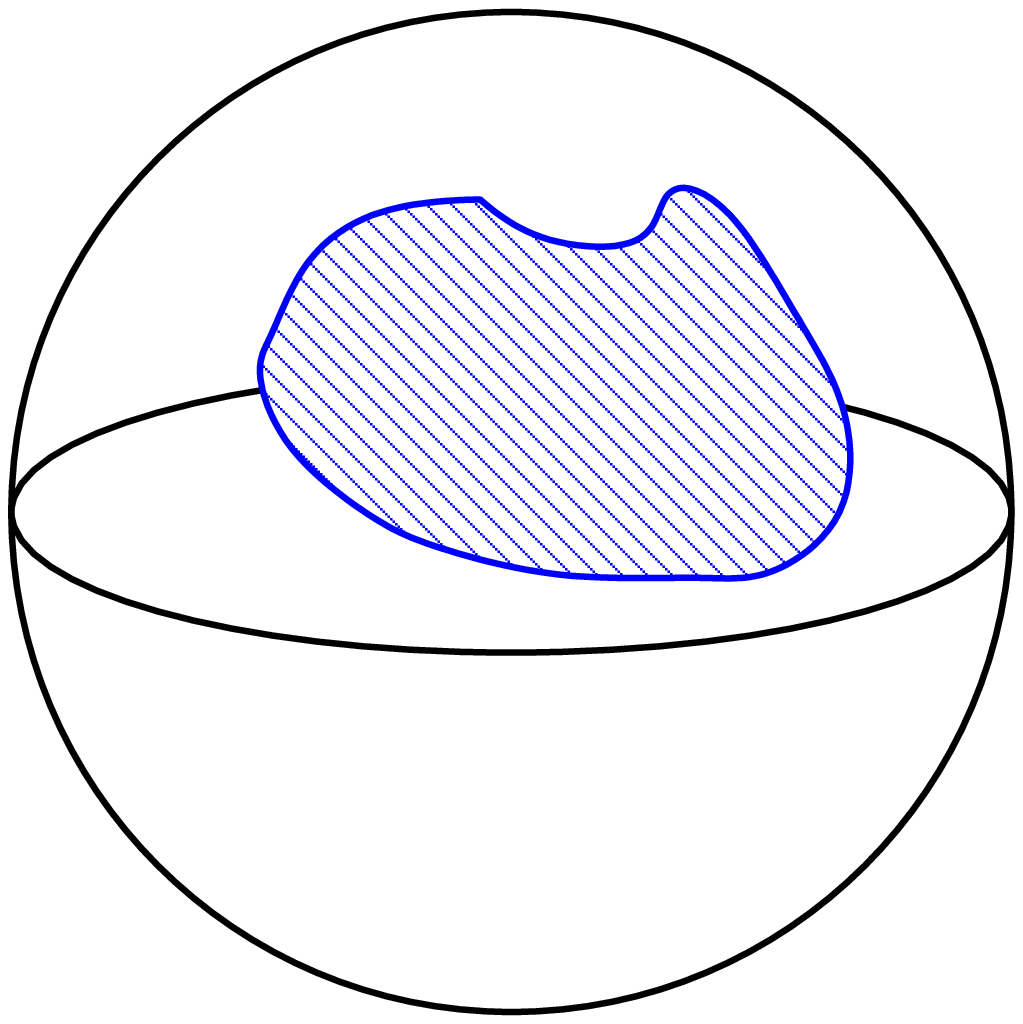}
\hspace{5mm}
\includegraphics[width=0.2\textwidth,height=0.2\textwidth,angle=0]{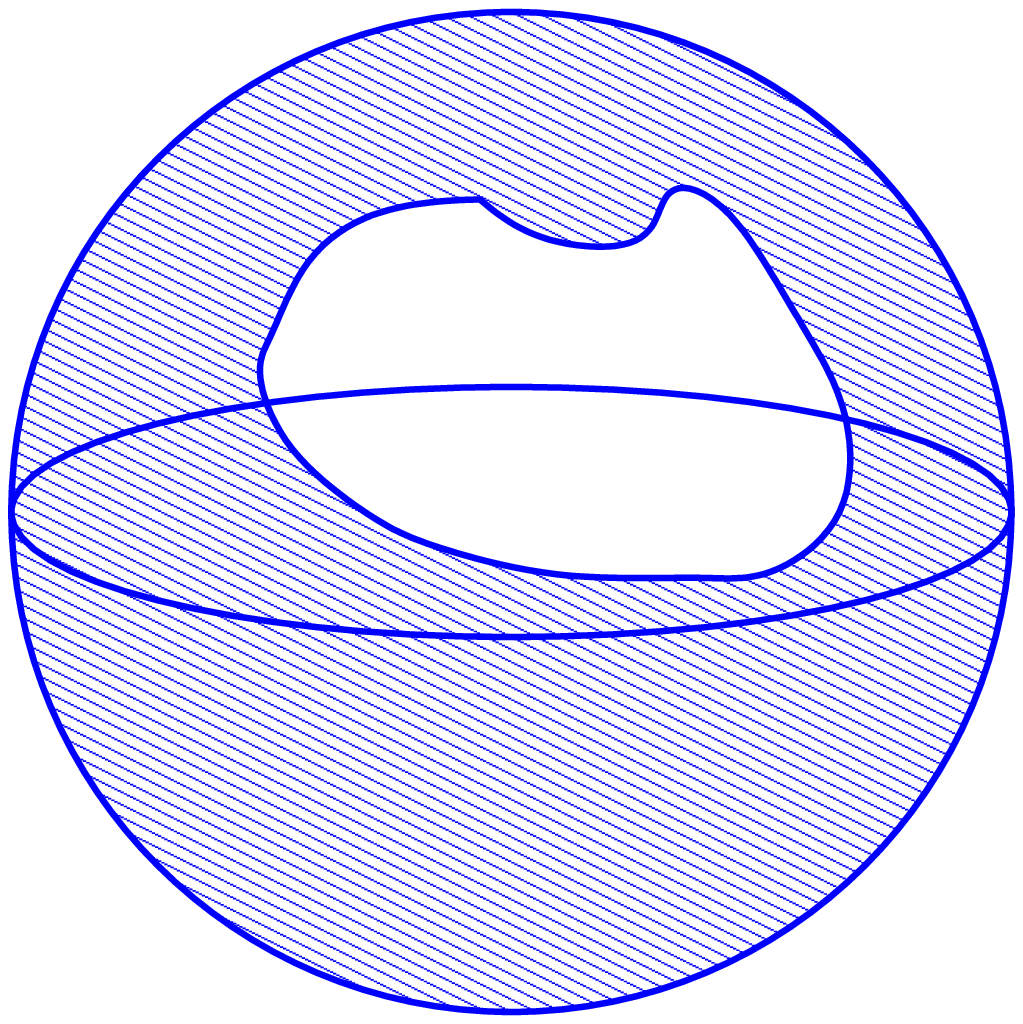}
\end{center}
\caption{Two possibilities for the 
surface whose boundary is the image of $\Phi$. 
The WZW action is given by the area of this surface.
}
\label{fig:1}       
\end{figure}

I describe here the analog of Witten's construction~\cite{Witten} on $SU(3)\times SU(3)/SU(3)\cong SU(3)$, but 
with the simpler topology of $SU(3)/SU(2)\cong S^5$.  This was outlined in 
\cite{Hill:2007nz} and details are presented in \cite{rjh}.~\footnote{
Including the $U(1)$ factor in $SU(3)\times U(1)/SU(2)\times U(1)$ is straightforward and
for simplicity it is not included here.
}
The topological complexity in the former case comes from the need to 
identify a five-sphere inside of $SU(3)$.   
Since the field space in the latter case {\it is} a five 
sphere, the WZW term for $SU(3)/SU(2)$ has a particularly simple construction.  
Note that just as reducing $SU(3)\times SU(3)/SU(3)$
to $SU(2)\times SU(2)/SU(2)$ describes the pion sector in the QCD chiral Lagrangian, 
so reducing $SU(3)\times U(1)/SU(2)\times U(1)$ to 
$SU(2)\times U(1)/U(1)$ describes the SM Higgs sector.~\footnote{
In fact additional interesting subtleties come up in this case - it turns
out that the reduction is only possible for an even number of ``colors''.
}

Fields on $SU(3)/SU(2)=S^5$ are represented by a vector 
$\Phi(x) = (\phi^1+i\phi^2, \phi^3+i\phi^4, \phi^5+i\phi^6)^T$ 
with 
\be
\Phi^\dagger \Phi = \sum_{i=1}^6 (\phi^i)^2 = 1 \,.
\ee   
What are the globally $SU(3)$ invariant
operators that we can write in terms of $\Phi$ ?  

Let us focus on the ungauged action. 
There is of course the ``ordinary'' class of operators, such as the kinetic energy 
term 
\be
{\cal L}_K = \partial_\mu \Phi^\dagger \partial^\mu \Phi + \dots \,. 
\ee
These operators are manifestly four-dimensional, local, and 
globally $SU(3)$ invariant.  

Another operator that is not so obvious is constructed as follows. 
Note that $x\mapsto \Phi(x)$ is a mapping from spacetime~\footnote{
To be precise, we consider Euclidean spacetime compactified onto $S^4$.
} into $S^5$.
Let the action 
be proportional to the area bounded by this 
mapping.   Explicitly, we have 
\be
\Gamma_{WZW}(\Phi) = {N\over \pi^2} \int_{M^5} \omega \,,
\ee
where $M^5$ is the surface with the image of spacetime
as its boundary, and 
\be 
\omega = {-i\over 8} \epsilon^{ABCDE} 
\Phi^\dagger \partial_A \Phi 
\partial_B \Phi^\dagger \partial_C \Phi 
\partial_D \Phi^\dagger \partial_E \Phi 
\,. 
\ee
is the volume element (=surface area) on the sphere. 
In fact, as shown in Figure~\ref{fig:1}, there are two different surfaces
with the image of spacetime as their boundary; for consistency, 
$e^{i\Gamma}$ should be independent of this choice.   Since the difference
is proportional to the total volume,~\footnote{Note the minus sign coming
from the relative orientation of the two surfaces.  
The total volume of $S^5$ is $\pi^3$. 
}
setting this difference equal to a multiple of $2\pi$ yields
the displayed quantization condition, with $N$ an even integer.  

Having constructed the WZW term, it is straightforward to check that
it is: 
(i) four dimensional (given $\Phi(x)$ defined on 4-d spacetime, 
we can compute $\Gamma(\phi)$); (ii) local (for a small change in $\Phi(x)$, 
the area defining the action changes by a small amount); and (iii) globally
$SU(3)$ invariant ($SU(3)$ acts as a subgroup of rotations on the sphere, and
the area is rotationally invariant).   
The action can be coupled to gauge fields for the $SU(3)$ generators, 
in which case it is still globally invariant, but generates an anomaly under 
local $SU(3)$ transformations.  This anomaly corresponds to that of $N$ left-handed 
fermions with global $SU(3)$ flavor symmetry. 

This construction can be formalized in terms of
the homotopy groups describing the topology of the 
sphere:~\footnote{
An equivalent statement can be made in terms of
de-Rham cohomology describing the classes of differential
forms that can be defined on the sphere~\cite{Weinberg}. 
}
$\pi_4(S^5)=0$, meaning that the construction is
possible (for a given $\Phi(x)$, there is a surface with 
the image of $\Phi(x)$ as its boundary); and 
$\pi_5(S^5)=\boldmath{Z}$, meaning that the construction is
nontrivial (the difference of the mappings in Figure~\ref{fig:1}
wraps the sphere nontrivially, and the action must be 
quantized).
This simple construction on the sphere carries over to 
more complicated spaces such as 
$SU(n)\times SU(n)/SU(n)=SU(n)$, $SU(n)/SO(n)$ and
$SU(2n)/Sp(2n)$. 

\section{Little Higgs models}
\label{sec:lh}

The starting point for so-called ``composite''~\cite{Kaplan:1983fs} and 
``Little''~\cite{nima,Schmaltz:2005ky} Higgs models 
is summarized by the formula for one-loop 
radiative corrections to pNGB masses.  
Suppose that we weakly 
gauge a collection of symmetry generators, 
\be
\Lambda = \Lambda_V + \Lambda_A \,,
\ee
where $\Lambda_V$ and $\Lambda_A$ are the unbroken 
and broken components of the generator.~\footnote{
The subscripts denote ``vector'' and ``axial'' components in
analogy to QCD-like symmetry-breaking patterns.
}
Then the mass-matrix for pNGB's is~\cite{massmatrix} 
\be\label{eq:masscor}
m_{ab}^2 = M^2 \sum_\Lambda 
{\rm Tr}\bigg\{ 
\left[ \Lambda_{V}, \left[ \Lambda_{V}, t^a_A \right] \right] t^b_A 
-
\left[ \Lambda_{A}, \left[ \Lambda_{A}, t^a_A \right] \right] t^b_A 
\bigg\} \,,
\ee
where $M^2$ is a nonperturbative 
mass scale set by underlying strong dynamics.  
If we suppose that an axial generator is gauged 
strongly enough so that $m^2 <0$ for the physical Higgs, then 
electroweak symmetry will be broken ``by vacuum misalignment''~\cite{Kaplan:1983fs}.   
Alternatively, 
suppose that the gauged generators are arranged so that, for the 
physical Higgs, $m^2=0$ through 
one-loop corrections.
Electroweak symmetry can then be
broken by higher-order loop corrections, and contributions from the 
top-quark sector~\cite{nima,Perelstein:2003wd}.
Assuming that the full electroweak symmetry breaking potential can be 
tuned or engineered in a plausible way, these models give a mechanism for 
a weakly coupled Higgs boson to leak down 
to the electroweak scale, along with
extra particles such as heavy 
partners of the top quark, and partners of the $SU(2)_L\times U(1)_Y$ gauge
bosons, that are involved in stabilising the Higgs mass against radiative corrections.  

The anomaly structure of such theories provides an important 
probe of the underyling UV completion physics.  The situation 
is analogous
to probing QCD 
if we only had access to sub-GeV experiments.  Anomaly physics enters in
two ways.  First, there are consistency conditions on the low-energy theory.  
For example, suppose we were able to deduce the electroweak gauge structure of
a complete first-generation standard model---the 
electron and its neutrino $(\nu_e, e)$, 
and the pions, coupled to $SU(2)_L \times U(1)_Y$.  
For a consistent gauge theory, we would find that the only possible value 
of the coefficient multiplying the WZW term is $N_c=3$.  
This provides an important clue to the UV completion, namely 
a theory of fundamental $SU(N_c=3)$ quarks.  
We would also know that interactions such as $\pi^0\to \gamma\gamma$ are not 
only possible in some UV completions, but required in any UV completion.   

Anomaly physics can also enter in a second way. 
Focusing just on the pseudoscalar + gauge boson sector,  
we can look for reactions 
such as $\pi^0\to \gamma\gamma$, and count $N_c=3$.~\footnote{
We of course need an independent measurement of $f_\pi$. 
}   
We then know that whatever other light fermion content exists 
must be consistent with this value.    

Let us look at a few simple examples to 
see how these arguments can be used to construct consistent 
Little Higgs models, and to constrain their UV completions. 

\subsection{ $SU(3)/SU(2)$ Little Higgs }

Consider first the implementation of the Little Higgs idea 
based on 
two copies of $SU(3)\times U(1) 
/ SU(2)\times U(1)$, introduced by
Kaplan and Schmaltz~\cite{Kaplan:2003uc,Schmaltz:2004de}.   
The model is described by two distinct ``condensate'' fields 
$\Phi_1$, $\Phi_2$, coupled to a single copy of 
$SU(3)_W \times U(1)_X$ gauge fields. 

When three fermion generations are present, and treated on the same
footing, there are uncancelled $SU(3)_W^3$, 
$SU(3)_W^2 \times U(1)_X$ and $U(1)_X^3$  
gauge anomalies (``Model 1'' of \cite{Schmaltz:2004de}).  
This is not necessarily a bad thing. 
In fact, if we let $N_1$ and $N_2$ be the coefficients of the WZW terms for
the $\Phi_1$ and $\Phi_2$ sectors~\cite{Hill:2007nz}, 
then we can enforce anomaly cancellation as long as 
\be
N_1 + N_2 = 12 \,,
\ee
or more generally $N_1+N_2 = 4 N_g$, with $N_g$ the number of 
generations. 
Without adding additional fields 
into the low-energy theory, there is no freedom to speculate
on the absence of the WZW term.   
Any candidate UV completion of this model 
must have exactly six ``colors'' per sector.~\footnote{
For example, $SU(3)/SU(2)$ can be embedded inside $SU(3)\times SU(3)/SU(3)$, 
and a potential UV completion consists 
of two triplets of ``techniquarks'' transforming under 
$\bm{6}$ and $\bar{\bm{6}}$ of $SU(6)$.~\cite{Hill:2007nz} 
} 
This is like the first application to the standard model discussed in the 
introduction to this section---before doing
any measurements, consistency places tight constraints on the form of the 
UV completion. 

By assigning different quantum numbers to the three generations, it is 
possible to cancel anomalies of the fermions amongst themselves
(``Model 2'' of \cite{Schmaltz:2004de}).  
In this case, we must have
\be
N_1 = - N_2 \,.
\ee
As we saw in Section~\ref{sec:simplest},
the ``number of colors'' must be even, so that we have the possibilities
$N_1 = -N_2 = 0, 2, 4, \dots\,.$  
An especially interesting scenario is where the two sectors 
are identical apart
from the chirality of the underlying condensates, represented by the
relative sign in front of the WZW term.
Then apart from Yukawa couplings to fermions, 
an exact exchange symmetry exists between the sectors, 
and is broken only by topological interactions.  
In terms of the physical
Higgs and $W$ boson, and the extra isosinglet scalar field $\eta$ and 
isodoublet vector field $C_\mu$ appearing in the model, 
we have~\cite{Hill:2007nz} 
\bea\label{simplestWZW}
&& \Gamma_{WZW} = \int d^4x\,\epsilon_{\mu\nu\rho\sigma}  \bigg\{ 
{-2N\over 8\pi^2\sqrt{3} F} \eta {\rm Tr}\big[ F_{W}^{\mu\nu} F_{W}^{\rho\sigma}\big] \nl
&& + {2N\over 16\pi^2 F} \big[ D^\mu H^\dagger F_W^{\nu\rho} C^\sigma 
+ h.c. 
\big]  + \dots \bigg\}  \,. 
\eea
This is like the second application to the standard model---
if we can identify and measure an anomaly-mediated interaction, 
we acquire a discrete and powerful probe of the UV completion. 

This example also illustrates a deficiency of the ``moose'' language of
links and sites that is sometimes used to describe Little Higgs models.  
In this language it appears
obvious that when the couplings in both sectors are identical, 
the reflection $\Phi_1 \leftrightarrow \Phi_2$ 
must be an exact symmetry.  
This led to the interesting proposal of 
an exact ``$T$'' parity~\cite{Cheng:2003ju} reflection symmetry, with 
implications for dark matter and missing energy collider 
signatures~\cite{Hubisz:2004ft,Carena:2006jx}.     
However, the parity is 
broken once anomalies are taken into account, since
$\Gamma_{WZW}$ in (\ref{simplestWZW}) is odd under this exchange.
These considerations apply to models invoking 
$T$ parity as a reflection symmetry 
in $[SU(3)\times SU(3)/SU(3)]^4$~\cite{Cheng:2003ju}.   
Similarly, the $T$-parity, or ``Goldstone boson parity'' appearing in 
the kinetic terms of symmetric-space cosets like 
$SU(3)\times SU(3)/SU(3)$, $SU(5)/SO(5)$ and $SU(6)/Sp(6)$ is 
violated by anomalies (in the QCD chiral Lagrangian
$\pi^0$ is odd under $T$ parity, the photon is even, 
yet $\pi^0\to \gamma\gamma$ is allowed!).
\footnote{
Some prospects for identifying an alternate ``$T$'' parity
were considered in \cite{Hill:2007zv}.  
This will generically require the existence or introduction 
of multiple condensate fields. 
}

\subsection{ $SU(5)/SO(5)$ Little Higgs } 

The usefulness of anomaly constraints apply more generally.
Here we discuss briefly two implementations of the Little Higgs
idea based on $SU(5)/SO(5)$.

In the so-called ``Littlest Higgs without $T$ parity'' 
model~\cite{ArkaniHamed:2002qy}, the 
extended-SM fermions couple only to one of the two gauged $SU(2)$
groups.   
This gauging can be arranged to be anomaly free, 
however two subtle problems arise.  
First, the fact that there is no 
left-over anomaly tells us that without extending the low-energy theory, 
the coefficient of the WZW term must be zero.
In the absence of additional fields, this rules out 
a technicolor-like UV completion~\cite{ArkaniHamed:2002qy} 
(either fundamental or composite fermions) 
that could have explained the origin of 
$SU(5)/SO(5)$ symmetry breaking. 
A more detailed study reveals another 
difficulty~\cite{Buras:2006wk}. 
The naive basis of operators in the theory specified by this gauging 
is not closed under renormalization.%
\footnote{
For example, both 
$\bar{\psi} \left( i\slash{\partial} + \slash{V} + \slash{A} \right)\psi$, 
and $\bar{\psi} U \left( i\slash{\partial} + \slash{V} - \slash{A} \right) U^\dagger \psi$ 
are gauge invariant kinetic terms for the fermions.  Here $V$ and $A$ denote 
broken and unbroken components of the gauge field, and $U$ is a unitary matrix of
pNGB's transforming as $U\to e^{i\epsilon} U e^{-i\tilde{\epsilon}}$, where the tilde changes the
sign of broken generators.  
}  
This 
limits the predictive power of the theory, even when restricted to low-energy
observables.

In the so-called ``Littlest Higgs with 
$T$ parity''~\cite{Cheng:2005as,Hubisz:2004ft}, 
apart from the breaking of $T$ parity by anomalies, 
we also run into problems of consistency.  The fermion content 
is anomalous 
(uncancelled $SU(2)^2 \times U(1)$ and $U(1)^3$ 
anomalies with the quantum numbers in \cite{Cheng:2005as,Hubisz:2004ft}). 
The form of these anomalies is such that they cannot be cancelled 
by the globally $SU(5)$ 
invariant WZW term, indicating that additional structure is required 
for gauge invariance. 
As a low energy theory, the model is inconsistent!

Concentrating just on the scalar + gauge sector, it is 
possible to look for anomaly-mediated interactions that could guide us to a more
complete model.  Some work along these 
lines is presented in \cite{Barger:2007df}.

\subsection{ Little Higgs Summary } 

It is essential in any low-energy effective theory to be able to write down
the most general operator consistent with an 
assumed field content and symmetry. 
This is especially important in bottom-up models such as the Little Higgs, 
where no particular UV completion is specified.  
Topological interactions represent one such class of operators. 
The anomaly structure encoded by these interactions 
is truly an IR probe of UV physics, 
providing consistency conditions on the low-energy theory, 
and constraints that any proposed UV completion 
must obey~\cite{Berezhiani:2005pb}.   

A perplexing folklore has developed in the Little Higgs literature,
whereby gauge invariance of the low-energy theory is considered optional.  
This is sometimes justified by the 
misleading argument that extra heavy fermions might 
exist that cancel anomalies.   
Of course, if the heavy fermions are truly heavy, 
they should be integrated out of the
low-energy theory, generating new operators
that maintain gauge invariance; 
if they are not
heavy, then they should be present in the 
low-energy theory, again maintaining gauge invariance.  
Such fermions will in general either be directly observable 
if they are light; 
or break global symmetries and affect the dynamics of 
electroweak symmetry breaking 
if they are heavy 
and transform under
an incomplete representation of the global symmetry group; 
or be a candidate to identify with underlying ``techniquarks'' of 
strong dynamics if they transform
under the complete representation of the global symmetry. 

Far from being a nuisance, anomalies and consistency conditions 
of the Little Higgs are one of the 
few handles we have to constrain the low-energy theory, and to probe
UV completion physics that is 
out of direct experimental 
reach. 

\section{Conclusion}
\label{sec:conclusion}

We can apply the technology developed for 
phenomenological applications in the standard model and 
Little Higgs theories to more general problems.   
For instance, in the study of formal large-$N$ equivalences between different fermion
field theories~\cite{Armoni:2003gp},
or between four-dimensional field theories and their conjectured holographic 
duals~\cite{Hill:2004uc}, 
the discrete nature of the WZW term can provide exact relations that 
are independent of $N$, or that are independent of wavefunction profiles in the
extra dimension.   
As a practical matter, these equivalences can be used as an efficient
calculational tool. 
For instance, the WZW term constructed directly on 
$SU(n)/SO(n)$~\cite{Auzzi:2006ns} 
can be derived immediately, including general gauge fields, from 
the WZW term for $SU(n)\times SU(n)/SU(n)$~\cite{Hill:2007nz}.    

\vskip 0.2in
\noindent
{\bf Acknowledgements}
\vskip 0.1in
\noindent
It is a pleasure to thank C.~Hill and J.~Harvey 
for collaboration on the topics in 
References~\cite{Hill:2007nz,Hill:2007zv,Harvey:2007rd}, 
especially those described here in Section~\ref{sec:weak} and 
Section~\ref{sec:lh}.  
Thanks also to N.~Arkani-Hamed, W.~Bardeen, H.C.~Cheng, 
J.~Hubisz, K.~Kong, E.~Lunghi and M.~Perelstein 
for interesting discussions on 
Little Higgs models. 
Research supported by the U.S.~Department of Energy  
grant DE-AC02-76CHO3000.

\end{document}